\documentclass[epj-spec]{svjour}
\usepackage{graphics}

\newcommand{\Hi}{\Delta_{\rm int}}

\begin{document}
\title{Spin relaxation times in disordered graphene }
\author{Daniel Huertas-Hernando\inst{1}\fnmsep\inst{2}\fnmsep\thanks{\email{danielhh@ntnu.no}} \and Francisco Guinea\inst{3} \and Arne Brataas\inst{1}\fnmsep\inst{2} }
\institute{Department of Physics, Norwegian University of Science and Technology, N-7491 Trondheim, Norway. \and Centre for Advanced Study, Drammensveien 78,  N-027 Oslo, Norway. \and Instituto de Ciencia de Materiales de Madrid, CSIC, Cantoblanco E28049 Madrid, Spain.}
%


\abstract{
We consider two mechanisms of spin relaxation in disordered graphene. i) Spin relaxation due to curvature spin orbit coupling caused by ripples. ii) Spin relaxation due to the interaction of the electronic spin with localized magnetic moments at the edges. We obtain analytical expressions for the spin relaxation times $\tau_{SO}$ and $\tau_{J}$ due to both mechanisms and estimate their values for realistic parameters of graphene samples. We obtain that spin relaxation originating from these mechanisms is very weak and spin coherence is expected in disordered graphene up to samples of length $\L \sim 1 \mu m$. 
} 
\maketitle
\section{Introduction}
\label{intro}

 Two dimensional (2D) graphene, a stable atomic layer of carbon atoms, 
remained for long ellusive among the known crystalline structures of carbon. 
Only recently, the experimental realization of stable, 
highly crystalline, single layer samples of
graphene  have been possible \cite{KN,Zhang}. 
Such experimental developments have generated a renewed interest 
in the field of two dimensional mesoscopic systems. The peculiar
electronic properties of the hexagonal lattice of carbon atoms makes 
graphene quite different from standard 2D 
semiconducting heterostructures samples. 

The unit cell of graphene is described by two inequivalent triangular 
sublattices $A$ and $B$ intercalated and there are two independent $k$-points, 
$K$ and $K'$, corresponding to the two inequivalent corners of the Brillouin zone
in reciprocal space.  The Fermi level is located at these K and K'
points and crosses the $\pi$ bands of graphene.
These two features provide an exotic fourfold degeneracy of the low
energy (spin-degenerate) states of graphene. These states can be described 
by two sets of two-dimensional chiral spinors which follow the massless
Dirac-Weyl equation in terms of a pseudo-spin degree of freedom which refers to the fourfold degeneracy mentioned. This Dirac-Weyl equation describes the electronic states of the system near the K and K' 
points where the Fermi level is located. At present, there is a large activity in the study of the dynamics of this pseudo-spin degree of freedom \cite{Proceedings}. 

Less attention has been given to the spin so far. The main interactions that could affect the spin degree of freedom in graphene seem to be the spin-orbit coupling and exchange interaction. It is not known to which extent magnetic impurities are present in actual graphene samples. Their effect seem small though, as noticed recently when investigating weak localization and universal conductance fluctuations in graphene\cite{Metal06}. 
Spin-orbit interaction in graphene is supposed to be weak, due to the low atomic number $Z=6$ of carbon. We have investigated recently the physics of the spin-orbit interaction in
graphene in detail\cite{DPA}. The effect of other interactions as the exchange interaction due to localized moments at the edges of the sample, has been discussed elsewhere\cite{SGV,VSG,PGCN}. 

\section{Spin Orbit }
\label{sec:1}
For the spin-orbit coupling in graphene  we have obtained an effective Hamiltonian for the $\pi$ bands, by second order perturbation theory, which is formally the same as the effective Hamiltonian obtained previously from group theoretical methods by Kane and Mele\cite{KM05}. We find for the coupling parameters charaterizing such a Hamiltonian that: i) The intrinsic interaction $\Hi \sim 10$mK is two orders of magnitude smaller than what was recently estimated by Kane and Mele\cite{KM05,DPA}. Moreover, we find that for typical values of the electric field as \emph{e.g.} used by Kane and Mele\cite{KM05}, the Rashba-type spin-orbit interaction due to an applied perpendicular electric field is $\Delta_{\cal E} \sim 70$mK.  So spin-orbit coupling for flat graphene is rather weak.  Graphene samples seem to have an undulating surface\cite{Metal06}. We have obtained that local curvature of the graphene plane induces another type of spin orbit coupling $\Delta_{\rm curv}$.  Our estimate for the typical observed ripples indicates that $\Delta_{\rm curv} \sim 0.2$K. It seems that curvature effects on the scale of the distance between neighbouring atoms could increase the strength of the spin-orbit coupling at least one order of magnitude with respect to that obtained for a flat surface. More importantly,  this type of  ``intrinsic'' coupling will be present in graphene as long as its surface is corrugated even if  $\cal E=$0 when $\Delta_{\cal E}=$0.

\begin{table}
\caption{Dependence on band structure parameters, curvature, and electric  field of the spin orbit couplings discussed in the text in the limit $V_1 \ll V_2$ (widely separated $\sigma$ bands). The parameters used are $\lambda \approx0.264$\AA, ${\cal E} \approx$ 50V/300nm, $\Delta = 12$meV,$V_{sp \sigma} \sim 4.2$eV, $V_{ss \sigma} \sim -3.63$eV, $V_{pp \sigma} \sim 5.38$eV and $V_{pp \pi} \sim -2.24$eV, $V_1=2.47$eV, $V_2=6.33$eV, $a=1.42$\AA,$~$ $l \sim 100$ \AA, $h \sim 10$ \AA $~$ and $R \sim 50-100$nm. See Ref. \cite{DPA} and references therein for details.}
\label{numerical}       




\begin{tabular}{lll}
\hline \hline
Intrinsic coupling: $\Hi$ & $\frac{3}{4} \frac{\Delta^2}{V_1} \left( \frac{V_1}{V_2} \right)^4$ &  0.01K\\ \hline
Rashba coupling (electric field ${\cal E}\approx 50V/300$nm): $\Delta_{\cal E}$ & $\frac{2 \sqrt{2}}{3} \frac{\Delta \lambda {e \cal E}}{ V_2}$& 0.07K\\ \hline
Curvature coupling: $\Delta_{\rm curv}$ & $\frac{\Delta ( V_{pp \sigma} - V_{pp \pi} ) }{V_1}
 \left( \frac{a}{R_1} + \frac{a}{R_2} \right)\left(\frac{V_1}{V_2} \right)^2  $&  0.2K\\ \hline
\end{tabular}

\end{table}

\subsection{Numerical estimate for Curvature Spin-Orbit.}
 $\Delta_{\rm curv}$ due to the corrugated nature of the samples, is the most important of the spin-orbit couplings. The ripples observed seem to be of several \AA$~$ height and a few tens nm laterally. It seems possible to identify  ripples of lateral size ranging  ~ 50nm -100nm in Ref. \cite{Metal06}. From these ripples we estimate $\Delta_{\rm curv} \sim $ 0.2 K\cite{DPA}. We have checked that this estimate, is in good agreement with recent data obtained for the band structure of the $\pi$- and $\sigma$-bands  \cite{Eli}, from which we obtain a value $\Delta_{\rm curv} \sim 0.15$K.

The effective Hamiltonian for graphene, including the spin-orbit coupling reads\cite{DPA}: 


\begin{equation}
{\cal H}_{T} = \int d^2 \vec{\bf r}  \Psi^\dag \left( -i \hbar v_F [\hat{\sigma}_x \hat{\partial}_x + \hat{\tau}_z \hat{\sigma}_y \hat{\partial}_y] + \Hi [\hat{\tau}_z \hat{\sigma}_z\hat{s}_z ] + \frac{\Delta_R }{2}[ \hat{\sigma}_x\hat{s}_y + \hat{\tau}_z \hat{\sigma}_y\hat{s}_x] \right) \Psi,
\label{Effective_H_SO}
\end{equation}
where $\hbar v_F =\sqrt{3} \gamma_o $a$ /2 $, a$~ \sim 2.46$ \AA $~$being the
lattice constant for graphene, $\gamma_o \sim 3$eV the McClure intralayer
coupling constant, $\Hi\sim
0.01$K, $\Delta_R =\Delta_{\cal E} +\Delta_{\rm curv}$ the Rashba-Curvature coupling (RCC), where $\Delta_{\cal E} \sim 0.07$K for ${\cal E} \approx$ 50V/300nm and $\Delta_{\rm curv} \sim 0.2$K. Table[\ref{numerical}] summarizes the main results obtained  for the effective spin-orbit couplings in Ref. \cite{DPA}. 

In the following, we consider ${\cal E} =0$, $\Delta_R =\Delta_{\rm curv}$ and $\Hi \ll \Delta_R$:

\begin{equation}
{\cal H}_{T} =  \int d^2 x \Psi^\dag(\vec{x})  \left( -i \hbar v_F [\hat{\sigma}_x\hat{\partial}_x +\hat{\tau}_z \hat{\sigma}_y\hat{\partial}_y] + \Delta_{R}[\hat{\sigma}_x\hat{s}_y +\hat{\tau}_z \hat{\sigma}_y\hat{s}_x] \right) \Psi(\vec{x})
\label{Effective_H_SO}
\end{equation}
We re-write ${\cal H}_{T}={\cal H}_{0}+{\cal H}_{R}$ where:
\begin{equation}
{\cal H}_{0} =  \int d^2 x \Psi^\dag(\vec{x}) \left( -i \hbar v_F [\hat{\sigma}_x\hat{\partial}_x +\hat{\tau}_z \hat{\sigma}_y\hat{\partial}_y] \right) \Psi(\vec{x}),
\label{Effective_H0}
\end{equation}
and
\begin{equation}
{\cal H}_{R}(\vec{x}) =  \int d^2 x \Psi^\dag (\vec{x}) \left( \Delta_{R}(\vec{x}) [\hat{\sigma}_x\hat{s}_y +\hat{\tau}_z \hat{\sigma}_y\hat{s}_x] \right) \Psi(\vec{x}) \equiv \int d^2 x \Delta_{R}(\vec{x})  \Psi^\dag (\vec{x}) \hat{\Gamma} \Psi(\vec{x}).
\label{Effective_Hdisorder}
\end{equation}
The coupling $\Delta_{R}(\vec{x})$ depends on the position of each ripple, as the corrugations depend on the position and the defined curvature may change sign across the sample randomly.  The average radius of curvature $R$ is zero and $\Delta_{R}(\vec{x})$ can be considered as a quenched Gaussian variable:

\begin{equation}
\left< \Delta_{R}(\vec{x}) \right>=0; ~~~ \left< \Delta_{R}(\vec{x}) \Delta_{R}(\vec{x^{'}}) \right> = \Delta_{R}^2 \delta^2(\vec{x}-\vec{x}^{'}) l^2,
\label{correlatx}
\end{equation} 
where $l$ provides the scale for both the average size of each ripple and the average distance between ripples and $<~>$ denotes ensemble averaging over different curved samples.

Following \cite{SGV} the self-energy due to this type of ``topological disorder''  reads in the SCBA: 
\begin{equation}
\Sigma_{\Gamma}= \left< \Delta_{R}(\vec{q}) \hat{\Gamma} G(\vec{p}+\vec{q} ,\epsilon) \hat{\Gamma} \Delta_{R}(-\vec{q}) \right>=\int \frac{ d^2 q}{(2 \pi)^2} \Delta_{R}(\vec{q}) \hat{\Gamma} G(\vec{p}+\vec{q} ,\epsilon) \hat{\Gamma} \Delta_{R}(-\vec{q}).
\label{Sigma_Gamma}
\end{equation}
Eq.(\ref{correlatx}) implies in momentum space:
\begin{equation}
\left< \Delta_{R}(\vec{q}) \Delta_{R}(-\vec{q}) \right> = \Delta_{R}^2 l^2,
\end{equation} 
and finally we have:

\begin{equation}
\Sigma_{\Gamma}= \Delta_{R}^2  l^2 Tr[ \int \frac{ d^2 k}{(2 \pi \hbar)^2} \hat{\Gamma}  G(\vec{k},\epsilon) \hat{\Gamma} ],
\end{equation}
where $Tr [~]$ denotes trace over all possible spin and pseudo-spin indexes and $ G(\vec{k},\epsilon)$ is the Green's function averaged over disorder.

We are interested in the components of the self-energy matrix which describe spin flip processes:
\begin{equation}
Im [\hat{s}_+ \Sigma_{\Gamma} \hat{s}_- ]  = \Delta_{R}^2 l^2 \frac{1}{\pi} \hat{s}_+ Im[ \int \frac{ d^2 k}{(2 \pi \hbar)^2}Tr[  \hat{\sigma}_+  G(\vec{k},\epsilon)\hat{\sigma}_+  ]] \hat{s}_-
\label{s+s-}
\end{equation}
and
\begin{equation}
Im [\hat{s}_- \Sigma_{\Gamma} \hat{s}_+ ]  = \Delta_{R}^2 l^2 \frac{1}{\pi} \hat{s}_- Im[ \int \frac{ d^2 k}{(2 \pi \hbar)^2}Tr[  \hat{\sigma}_-  G(\vec{k},\epsilon)\hat{\sigma}_-  ]] \hat{s}_+,
\label{s-s+}
\end{equation}
where now $Tr [~]$ is only trace over pseudo-spin indexes. Note that the processes that provide spin flip, Eq.(\ref{s+s-}) and Eq.(\ref{s-s+}), are contributions in which $K$ and $K'$ are mixed by disorder. The only finite contributions to spin flip are contributions \emph{e.g.} where $\tau_z= +1$ for the first $\Gamma$ and $\tau_z=-1$ for the second $\Gamma$ in Eq.(\ref{Sigma_Gamma}). A process $\uparrow \rightarrow \downarrow$ requires $K \rightarrow K'$ and  for  $\downarrow \rightarrow \uparrow$ requires $K' \rightarrow K$.
 Summing up both contributions, we finally get for the imaginary part of 
$\Sigma_{curv}= \hat{s}_+\Sigma_{\Gamma} \hat{s}_- +  \hat{s}_- \Sigma_{\Gamma} \hat{s}_+$:
\begin{equation}
Im [\Sigma_{curv} ]  = \Delta_{R}^2 l^2 \frac{1}{\pi} Im[ \int \frac{ d^2 k}{(2 \pi \hbar)^2}Tr[  G(\vec{k},\epsilon)]~ ] =  \Delta_{R}^2 l^2 \nu(\epsilon).
\label{Sigmacurv}
\end{equation}
Note that there is no extra valley degeneracy and in this case spin and valley degeneracy are coupled. This means that $\nu(\epsilon)$ in Eq.(\ref{Sigmacurv}) is $1/2$ of the total density of states for a disordered graphene sample $\nu_T(\epsilon)=g_s g_v |\epsilon|/ (2 \pi (\hbar v_F)^{2})$, where $g_s= g_v=2$ are the spin and valley degeneracies respectively.
At zero applied perpendicular electric field, $\Delta_{R}=\Delta_{\rm curv}$ and we obtain the spin relaxation time due to curvature spin-orbit coupling as:
\begin{equation}
\frac{\hbar}{\tau_{SO}}= Im [\hat{s}_+ \Sigma_{\Gamma} \hat{s}_-] = \Delta_{\rm curv}^2  l^2 \nu(\epsilon). \end{equation}
In clean graphene the ``spin-valley'' density of states for energies close to the Dirac point, $\epsilon \simeq 0$, reads:
\begin{equation}
\nu(\epsilon)= 2 \times 2 \frac{|\epsilon|}{2 \pi (\hbar v_F)^{2}} \times \frac{1}{2}=\frac{|\epsilon|}{\pi (\hbar v_F)^{2}}.
\end{equation}
The presence of vacancies or dislocations leads to a finite density states at the Dirac point, $\epsilon = 0~$\cite{PGCN}, which in our ``spin-valley'' case reads:
\begin{equation}
\nu(\epsilon = 0 ) \simeq \frac{1}{\hbar v_F \bf {l}}. 
\end{equation}
$\bf {l}$ is the mean free path associated with the dislocations or vancancies in the system\cite{PGCN}.
We assume that due to dislocations or vacancies, there is finite intervalley scattering  in the sample at length scales of the order of $\bf {l}$. Also if the graphene sample is tightly coupled to the insulating SiO$_2$ substrate, this generates atomically sharp scatterers which lead to intervalley scattering\cite{PRLMcCann}. 
Using $\hbar v_F = \frac {3}{2} t a \sim 5.3 eV \AA$, where  $t$ is the hopping between neighboring atoms in the graphene lattice, ${\bf l} = 10^{2}a$, where $a$ is the lattice constant of graphene, $a=2.46 \AA$, $l=100 \AA$ and $ \Delta_{\rm curv} \sim 0.2 K \sim 1.742 \times 10 ^{-5} eV$, we finally have for the spin-relaxation time  close to the Dirac points $K, K^{'}$:

\begin{equation}
\tau_{SO}= \frac{\hbar ^{2} v_F \bf{l}} { \Delta_{\rm curv}^2 l^{2}} 
\sim 0.282 \times 10 ^{-6} s. 
\end{equation}


\section{Localized moments}
Lattice distorsions and sample edges in graphene can induce localized states at the Fermi energy, leading to the existence of local moments\cite{VSG}. The RKKY interaction between these moments is always ferromagnetic due to the semimetallic properties of graphene. Instability of the paramagnetic phase due to electron-electron interaction has been also proposed as mechanisms leading to ferromagnetism in graphene\cite{PGCN}. Also, spin-flip  for spin splitted edge channels in the QH regime has been estimated to be rather  weak\cite{AbaninLevitov}. 

The interaction between the electronic spin and the localized magnetic moments is given by Heisenberg Hamiltonian:
\begin{equation}
{\cal H}_{J} = J \hat{\bf {S}} \cdot \hat{s}
\end{equation}
where
\begin{equation}
J=\frac{t^2}{U} \sim U 
\end{equation}
where $ t \sim U $ because for graphene $e^2/\hbar v_F \equiv (e^2)/ (W a) \equiv U/W \sim 1$. 

It is straightforward to obtain the spin-relaxation time due to the interaction between the electronic spins and the localized moments:

\begin{equation}
\frac{\hbar}{\tau_{J}}\sim S(S+1) t^2 \frac{\bar{N} a^2}{\hbar v_F \bf{l}}.
\end{equation}
 $\bar{N}$ denotes the total probability for an electronic quasiparticle to reach the edges and interact with the localized moment. Using $\bar{N}=N f (a/L)^2$, where $N$ is the fraction of localized moments per edge and $f$ is the fraction of edge which is a zigzag where localized states occur,  we finally have:

\begin{equation}
\tau_{J} =\hbar \left(\frac{3}{2}\right)^{2} \frac{\bf{l}}{\hbar v_F} \frac{1}{S(S+1) N f}\left(\frac{L}{a}\right)^{2}. 
\end{equation}
Taking $N=0.3$, \emph{i.e.} there is a localized moment every three atoms of a zizag edge, $f=0.2$ as the average fraction of a sample edge which is of zigzag type  and for the  sample size $L=100 \mu$m, we have 
\begin{equation}
\tau_{J} \sim 0.16 \times 10 ^{-6} s.
\end{equation}

\section{Diffusion coefficient in graphene}
\label{sec:2}

From the mobility of the graphene samples $\mu = \sigma/(e n)$, $n$ being the electron(hole) density and $e$ the electronic charge, and using the Einstein relation $\sigma = e^{2} \nu(\epsilon_F) D$, we obtain for the diffusion coefficient:

\begin{equation}
D \sim \frac{\mu \hbar v_F {\bf l} n }{e}. 
\end{equation}
For $\mu \sim 10^{4} cm^{2}/(V s)$\cite{KN}, $n= 10^{11} cm^{-2}$ \cite{CNGph}, we obtain a very rough estimate for the diffusion coefficient:

\begin{equation}
D \sim  2 \times 10 ^{-5} m^{2} s^{-1}.
\end{equation}
Under the assumption that transport in disordered graphene is diffusive, we obtain the following  order of magnitude for the spin-flip lengths:

\begin{equation}
\lambda_{SO}=\sqrt{D \tau_{SO}} \sim \lambda_{J}=\sqrt{D \tau_{J}} \sim \sqrt{0.2 ~ 10 ^{-6} \times 2 ~ 10 ^{-5} } \sim 2 \mu m. 
\end{equation}

\section{Conclusions}
We have investigated  two mechanisms of spin relaxation in disordered graphene: i) Spin-flip due to the spin-orbit interaction caused by ripples in the surface of graphene and ii) Spin-flip due to the interaction of the spin with magnetic moments localized at the sample edges.
We have found out that spin-relaxation from these mechanisms is very weak with a relaxation time $\tau_{SO(J)}\sim 0.2 \times 10 ^{-6}s$. By estimating the diffusion coefficient in graphene samples, we obtain that the spin-flip length due to spin-orbit and/or interaction with localized moments $\lambda_{SO(J)}$ is of the orther of $\sim 1 \mu$m. So spin relaxation should start being important if the sample size $L$ becomes of the other of  $L \sim 1 \mu$m. This shows that spin coherence in graphene can be preserved for long distances. These results support perspectives for spintronics applications of graphene.

We thank Alberto Cortijo, Maria A. H. Vozmediano and Bart van Wees for valuable discussions. D.H-H and A.B. acknowledge funding from the Research Council of Norway, through grants no 162742/v00, 1585181/431 and 1158547/431.  F. G. acknowledges funding from MEC (Spain) through grant FIS2005-05478-C02-01 and the European Union Contract 12881 (NEST) and the Comunidad de Madrid, 
through the program CITECNOMIK, CM2006-S-0505-ESP-0337.

%

%


\begin{thebibliography}{}
\bibitem{Proceedings} A good review of the present activity should be these Proceedings.
\end{thebibliography}
\end{document}